# Aligning the Western Balkans power sectors with the European Green Deal

Emir Fejzić[1,*], email: fejzic@kth.se, ORCID iD: https://orcid.org/0000-0002-2489-8455
Taco Niet[2], email: taco_niet@sfu.ca, ORCID iD: https://orcid.org/0000-0003-0266-2705
Cameron Wade[3], email: cameron@sutubra.ca, ORCID iD: https://orcid.org/0000-0003-2680-6881
Will Usher[1], email: wusher@kth.se, ORCID iD: https://orcid.org/0000-0001-9367-1791

[1] KTH, School of Industrial Engineering and Management (ITM), Energy Technology, Energy Systems, Brinellvägen 68
SE-100 44 Stockholm, Sweden
[2] Simon Fraser University, School of Sustainable Energy Engineering, 10285 University Drive, Surrey, BC V3T 4B7, Canada.
[3] Sutubra Research Inc. 5608 Morris St, Halifax, Nova Scotia. B3J 1C2, Canada

## Abstract

Located in Southern Europe, the Drina River Basin is shared between Bosnia and Herzegovina, Montenegro, and Serbia. The power sectors of the three countries have an exceptionally high dependence on coal for power generation. In this paper, we analyse different development pathways for achieving climate neutrality in these countries and explore the potential of variable renewable energy (VRE) and its role in power sector decarbonization. We investigate whether hydro and non-hydro renewables can enable a net-zero transition by 2050 and how VRE might affect the hydropower cascade shared by the three countries. The Open-Source Energy Modelling System (OSeMOSYS) was used to develop a model representation of the countries' power sectors. Findings show that the renewable potential of the countries is a significant 94.4 GW. This potential is 68% higher than previous assessments have shown. Under an Emission Limit scenario assuming net zero by 2050, 17% of this VRE potential is utilized to support the decarbonization of the power sectors. Additional findings show a limited impact of VRE technologies on total power generation output from the hydropower cascade. However, increased solar deployment shifts the operation of the cascade to increased short-term balancing, moving from baseload to more responsive power generation patterns. Prolonged use of thermal power plants is observed under scenarios assuming high wholesale electricity prices, leading to increased emissions. Results from scenarios with low cost of electricity trade suggest power sector developments that lead to decreased energy security.

# 1  Introduction

The impacts of climate change are observed all around the world. The severity and frequency of extreme climate are driven by anthropogenic greenhouse gas (GHG) emissions. [1]. To mitigate climate-induced impacts on the environment and society, it is imperative to reduce our GHG emissions. Recent figures show that globally, the share of carbon dioxide (CO2) accounts for 64% of total GHG emissions [2]. Out of the 36.3 gigatonnes (Gt) of CO2 emissions from energy-related activities, 10.5 Gt come from coal-fired thermal power plants (CFTPP) [3]. This represents 29 % of total energy-related $CO_2$ emissions. Given these statistics, traditional coal power plants must be replaced with low-emitting renewable energy sources. One region with an exceptionally high dependence on coal for power generation is The Western Balkans (WB). Coal shares in the power sectors of this region range from 55% in North Macedonia to 97% in Kosovo[1] [4] Therefore, low-carbon development pathways must be explored to accomplish the climate and environmental objectives of the WB countries, including the Sustainable Development Goals (SDGs) in the 2030 Agenda and alignment with the European Green Deal under the Sofia Declaration.

Renewable energy sources provide a cleaner alternative to the region's reliance on coal. Among the identified sources of renewable energy in the WB is the Drina River basin (DRB). It boasts a significant hydropower potential, of which 60% remains untapped [5]. The basin is shared by three countries: Bosnia and Herzegovina, Montenegro, and Serbia. These countries have applied to join the European Union (EU) and pledged Nationally Determined Contributions (NDC) as part of the 2030 Agenda. However, for the DRB countries to achieve decarbonization and to align with climate policies and objectives implemented at the EU and global levels, they must incorporate renewable resources other than hydropower into their energy mix. Consequently, the potential for variable renewable energy (VRE) in DRB countries must be assessed.

Current literature shows a lack of consistency regarding VRE potential estimates for the DRB countries. Hrnčić et al. [6] investigated the possibility of achieving a 100% renewable energy system in Montenegro. The wind power potential assumed in the study was 400 MW, and it referred to the Energy Development Strategy of Montenegro until 2030 [7] and Vujadinović et al. [8]. All three studies claim 400 MW of technical wind power potential in Montenegro, a figure taken from an assessment conducted by the Italian Ministry for the Environment, Land and Sea in 2007 [9]. Wind turbines have developed rapidly since 2007. Wiser et al. [10] found in 2012 that the land area in the US where wind power plants could achieve capacity factors (CFs) of 35% or higher increased by 260% when using turbines designed for the International Electrotechnical Commission (IEC) Class III wind conditions compared to turbines from the 2002-2003 era. Moreover, a report published by the International Renewable Energy Agency (IRENA) in 2017 [11] assessed the technical potential of VRE in the DRB countries to be 56.3 gigawatts (GW). The wind power potential for Montenegro is, according to [11] close to 2.9 GW, a much greater potential than earlier estimates of 400 MW.

---

[1] This designation is without prejudice to positions on status and is in line with UNSCR 1244 and the ICJ Opinion on the Kosovo Declaration of Independence.

Studies including Hrnčić et al. [6] and Husika et al. [12] use energy models to investigate potential development pathways for the power sectors of Bosnia and Herzegovina and Montenegro. Other recent publications confirm the lack of data on renewable energy potentials in Bosnia and Herzegovina [13]. No studies have included decarbonization pathways by 2050 for all DRB countries and their shared hydropower potential within the DRB. An existing study by Almulla et al. [14] uses OSeMOSYS to investigate the benefits associated with optimised production and increased cooperation between hydropower plants (HPP) in the DRB, including the impacts of energy efficiency measures. However, since the model, projections, and comparison periods used in [14] differ, the overall methodology and approach can be differentiated from the approach utilized in this study.

This paper aims to fill the identified research gaps by investigating decarbonization pathways for the DRB countries' climate neutrality by 2050. In addition, we estimate the power potential of VRE technologies within the DRB countries, which can help facilitate this transition away from coal-based power generation.

In this paper, we aim to answer the following research questions (RQs):
- What is the VRE potential in the DRB countries, and what role can it play in decarbonizing the power sectors?
- Are the resource potentials of hydro and non-hydro renewables in the DRB countries enough to support the transition to net zero by 2050?
- What is the impact of VRE on the existing hydropower cascade in terms of power generation and cost competitiveness?

In section 2, this paper provides a background of the DRB countries and identifies research gaps. It then describes the methodology used in section 3. We present the choice of modelling tool, the temporal and geographical dimensions of the study, and the operational constraints. The results and discussion are presented in sections 4 and 5, respectively. Section 6 concludes the work, followed by section 7, which identifies limitations and future research.

## 2 Background

The Western Balkan (WB6) countries of Albania, Bosnia and Herzegovina, Kosovo, Montenegro, North Macedonia, and Serbia are the only Southern European[2] countries that are not yet part of the European Union (EU) [16]. Fig. 1 shows the DRB area, which covers most of the cross-border area between the DRB countries [17]. The basin area is 20,320 km² and corresponds to 14% of the total land area of the DRB countries [18].

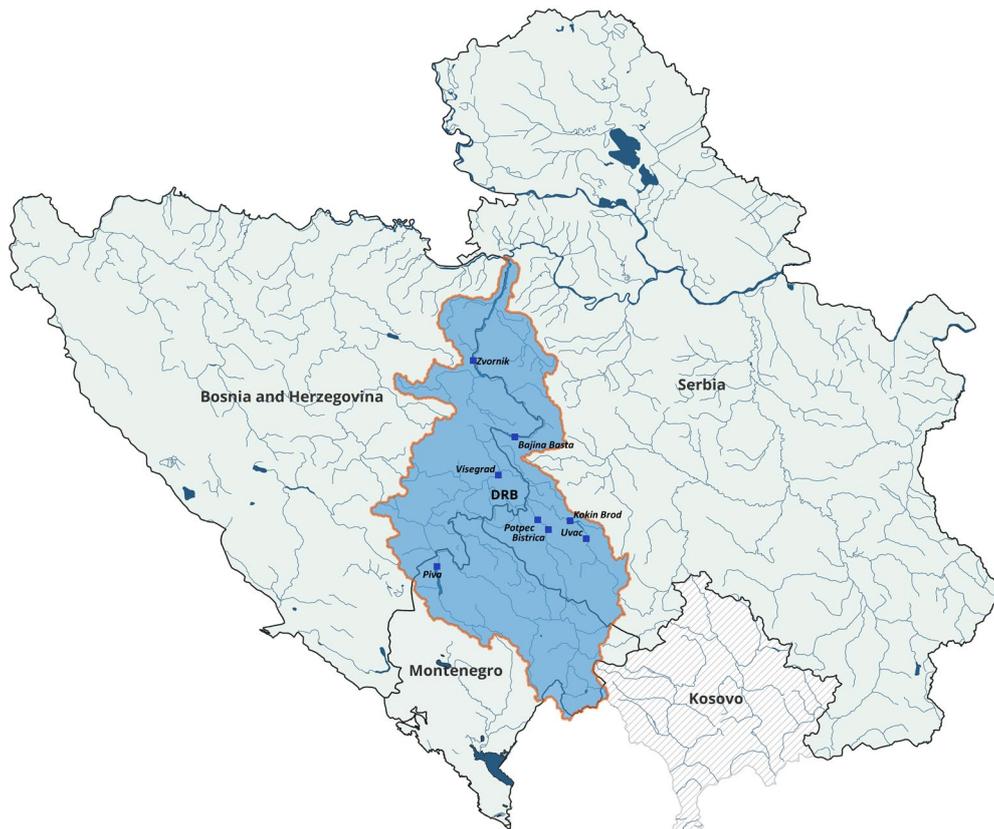

**Fig. 1**. The Drina River Basin (outlined in orange) modelled hydropower plants within the basin (blue squares) and the DRB countries.

### 2.1 Overview of the countries

In 2021, the population of the DRB countries was 10.7 million [19]. Serbia is the largest country with 6.8 million residents, followed by Bosnia and Herzegovina and Montenegro, with 3.3 and 0.6 million, respectively. Out of 867 thousand people living within the DRB, 50% reside in Bosnia and Herzegovina, 33% in Serbia, and 17% in Montenegro [20]. Montenegro and Serbia have a per capita GDP of approximately nine thousand USD, while Bosnia and Herzegovina has seven [21]. The average GDP per capita in the EU is 38 thousand [22], significantly higher than in the DRB countries.

Similarly, power consumption per capita is lower than the EU average. In Bosnia and Herzegovina, Montenegro, and Serbia, the consumption is 3.3 MWh [23], 5.7 MWh [24], and 4.1 MWh [25], respectively. By contrast, the EU-27 consumption is 6.4 MWh [26, 27].

---

[2] According to the United Nations classification and definition of regions [15] UN ESA. "Classification and definition of regions." https://esa.un.org/MigFlows/Definition%20of%20regions.pdf (accessed 2022.09.15, 2022).

The DRB countries depend heavily on coal-fired thermal power plants (CFTPP). Between 2014 and 2018, the share of CFTPP in the power supply was 60-70% in Bosnia and Herzegovina [28] and Serbia [29], reaching 40% in Montenegro [30]. This indicates that reliance on coal is significantly higher than the 20% share of coal in the EU power sectors [31]. Realizing this reliance on coal by the DRB countries, the European Commission (EC) launched the *Initiative for coal regions in transition* in December 2020 [32]. This initiative aims to assist the Western Balkans, including the DRB countries, in their transition from coal to carbon-neutral economies.

Coal-fired power generation has other adverse effects beyond those related to climate change. Estimates indicate 880 deaths from air pollutants in 2020 alone, resulting from exceeding the National Emissions Reduction Plans (NERP) greenhouse gas ceilings by CFTPPs in Bosnia and Herzegovina and Serbia. This includes 235 deaths due to exports from the countries to the European Union. Health costs from overshooting GHG emissions in the WB countries are estimated to be between 6-12 billion euros in 2020 [33]. Reducing coal use will help reach the climate goals and improve air quality, preventing chronic illnesses and premature deaths associated with PM, SO2, and NOx pollutants.

## 2.2 Aligning with EU climate goals

The DRB countries are candidates for EU accession, with Bosnia and Herzegovina obtaining the status as of December 2022 [34]. As EU member state candidates, the countries have pledged to align with the EU Climate Law and the EU Emissions Trading Scheme (EU-ETS) and to increase their share of renewable energy sources by the signing of the Sofia Declaration in 2020 [35]. The alignment with EU policy entails that by the time the countries become EU member states, their climate and energy policies must align with the European Green Deal, implying a 55% emission reduction by 2030 compared to 1990 levels and achieving climate neutrality by 2050 [36].

## 2.3 Nationally Determined Contributions as part of the Paris Agreement

The DRB countries have submitted their NDCs to the United Nations Framework Convention on Climate Change as part of their efforts to combat climate change. Each country has since submitted updated versions of its NDCs, increasing the ambitions compared to the first submission. Table 1 summarizes the pledged decreases in GHG emissions submitted in their NDCs. In addition to the 2030 goals, Bosnia and Herzegovina is committed to reducing emissions by 61.7% unconditionally and 65.6% conditionally by 2050 compared to 1990 levels. As stated, the contributions do not align with the EU targets, and ambitions must be raised to reach climate neutrality by 2050.

**Table 1.** Pledged emission reductions by the DRB countries by 2030 relative to 1990 levels.

| Country | First NDC [%] By 2030 | Updated NDC [%] By 2030 |
|---|---|---|
| Bosnia and Herzegovina | 18 and 20% (conditional and unconditional) | -36.8 and -33.2 (conditional and unconditional) |
| Montenegro | -30% | -35% |
| Serbia | -9.80% | -33.30% |

### 2.4 Renewable resource potentials

Located in Southern Europe, the DRB countries have a higher photovoltaic power potential compared to the northern and central parts of Europe [37]. Earlier studies for the Central and South East Europe region have identified a significant potential for renewable energy technologies [11, 38]. Currently, these potentials are largely untapped. The literature does not agree on the potential estimates of renewable energy sources within the DRB countries. To do so, it is critical to employ a consistent methodology across the three countries using the latest high-resolution geospatial and temporal data.

## 3 Methodology

In this section we describe the structure of the energy system model of the DRB countries. In addition, we describe and justify the choice of methods to assess wind and solar potential in the DRB countries. This includes the selection of the modelling framework, data, and methodology for determining the power potentials from VRE sources. Next, we present the clustering approach to manage computational effort, which retains critical temporal details across electricity demand and VRE sources. In addition to the model being open source, the clustering approach ensures that it is also accessible given a reduced computational effort. Finally, we present the scenario analysis.

### 3.1 OSeMOSYS and model setup

To answer the three research questions posed in Section 1, namely what the potential of VRE in the DRB countries is, if they are sufficient to support the transition to net-zero by 2050, and what their impact is on the existing HPP cascade, the created energy model must possess certain qualities. It should be geospatially explicit, allowing for assessment of VRE potential within DRB countries. The model must account for daily and seasonal climate and power demand variations while minimizing computation effort. Modelling HPP cascades on a per-power plant basis is required to assess how changes to the system-wide infrastructure affect these cascades. It must provide insight into future development pathways for the power sector by presenting a long-term power system expansion. A detailed description of the OSeMOSYS framework used to represent the hydropower cascade within the Drina River basin, together with the interconnected energy systems of the Drina River basin countries, is presented in this section.

We created the model using the OSeMOSYS framework [39, 40]. The primary use of OSeMOSYS is for long-term energy planning based on systems optimization. It does not require proprietary software or commercial programming languages and solvers. It is, for this reason, a preferable option compared to long-established models such as MARKAL/TIMES [41], MESSAGE [42], and PRIMES [43] to name a few, as it does not require upfront costs. The OSeMOSYS framework consists of seven blocks. These blocks are defined as the objective function, costs, storage, capacity adequacy, energy balance, constraints, and emissions. The objective function is the total discounted energy system cost in this case. It is based on provided energy carrier demands. Costs include capital investment costs, operating costs, and salvage values, among others [39]. Constraints include a reserve margin constraint, which, in the case of this analysis, is set to be 20%. The reserve margin is based on the fact that power generating companies and transmission companies must maintain a capacity to generate and transmit electricity exceeding normal capacity by 10-20% [44]. While multiple emissions can be attributed to power-generating technologies or resource extractions, we consider $CO_2$ emissions in this study. Costs within the model are discounted at a global discount rate of 5%.

As a basis for developing the model, we collected data from the literature, stakeholder engagements with representatives from the DRB, transmission system operators (TSOs), and directly from their respective power utilities. The data types collected include power demands, installed power generating capacities, fixed and variable costs of power plants, resource potentials and fuel costs, and cross-border transmission capacities, to name a few. All data used for the model creation, including the data files used to run each scenario and scripts used for assessing the VRE power potential, can be found in the following [Github](#) repository and [Zenodo](#) deposit.

### 3.2 VRE Characterization

The following section describes the methodology behind the model's characterization of solar and wind power potentials. We provide details on the approach taken to assess the resource availability of VRE and its power generation potential.

#### 3.2.1 Resource availability

For the VRE potential assessment, we calculate each country's total land (km2) eligible for VRE development. We assume that wind and solar can be developed on the following land use types in the CORINE Land Cover (CLC) database: natural grasslands, moors and heathland, sclerophyllous vegetation, transnational woodland-shrub, bare rocks, sparsely vegetated areas, and burnt areas. The resulting land availability representation is shown in Fig. 2.

We refer to the squares representing the eligible land fractions in Fig. 2 as grid cells. Each grid cell has a 30 x 30 km resolution. To obtain hourly time series data on VRE power generation potentials in each of the grid cells, we used Atlite [45], a Python package for calculating renewable power potentials and time series. As Atlite utilizes the ERA5 dataset, we chose the 30 x 30 km resolution over the higher-resolution CLC data.

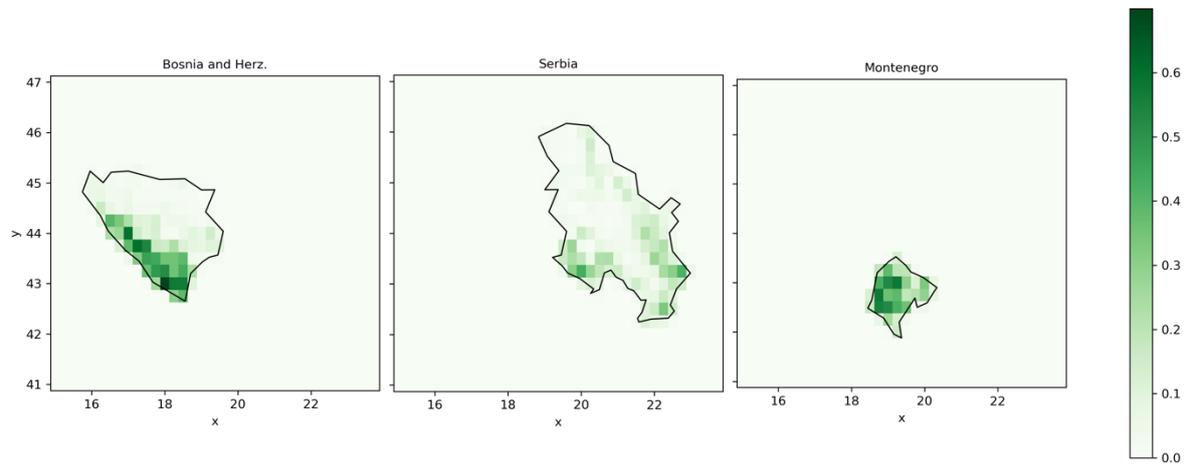

**Fig. 2.** Eligible land fraction for wind and solar power plants (excluding agricultural land) for the DRB countries. The axis labels show longitude (y-axis) and latitude (x-axis). The resolution is a 30km grid.

The total potential for wind and solar development expressed in terms of capacity (MW) is calculated by multiplying the total eligible land by an area-specific maximum installable capacity of 1.7 MW/km2. The maximum installable capacity is based on [46] and is used in similar studies [47]. The number consists of a technical potential density for installable wind generation capacity, corresponding to 10 MW/km2, and a fraction of 0.17, including the consideration for public acceptance and competing land use, extreme slopes, and unfavourable terrain. The more precise the location analysis is, the higher the area-specific installable capacity number can be used. To determine the potential capacity of wind and solar power in each DRB country, we multiply the capacity per square kilometre by the total eligible land area shown in Fig. 2.

3.2.2   VRE generation potentials

In addition to maximum resource potentials, the OSeMOSYS model requires time series values for the production potentials. These are used to calculate the CFs for each wind and solar technology included in the model.

As land availability affects the distribution of VRE resources, it is necessary to consider the weather variations according to the location of the VRE instalments. We use Atlite [45] to estimate the hourly wind and solar production potential in each grid cell. Atlite retrieves global historical weather data and converts it into power generation potentials and time series for VRE technologies like wind and solar power. The data used has an hourly temporal resolution obtained from the fifth-generation European Centre for Medium-Range Weather Forecasts (ECMWF) atmospheric reanalysis of the global climate (ERA5 dataset). To get a representative year, we first calculated the average hourly power output for wind and solar from the five years of 2017-2021. Next, we selected the year that best represented the average, in this case, 2020. We chose to use a historical weather year instead of an average year since the average weather year increases the lower extremes and decreases the higher extremes.

To translate the ERA5 weather data to capacity factors, we use the following technology assumptions: for wind, we chose a Siemens SWT-2.3-108 turbine, with a rated power of 2.3 MW and a hub height of 100m [48]. It has a cut-in wind speed of 3 m/s and a cut-out speed of 25 m/s. The power curve of the selected wind turbine type is shown in Fig. 3. Cadmium telluride (CdTe) photovoltaics (PV) solar panels with the orientation 'latitude optimal' were selected in Atlite for solar power. The CdTe panel characteristics are provided by [49]. No tracking was included for the solar panels, as Atlite did not include tracking when this analysis was done.

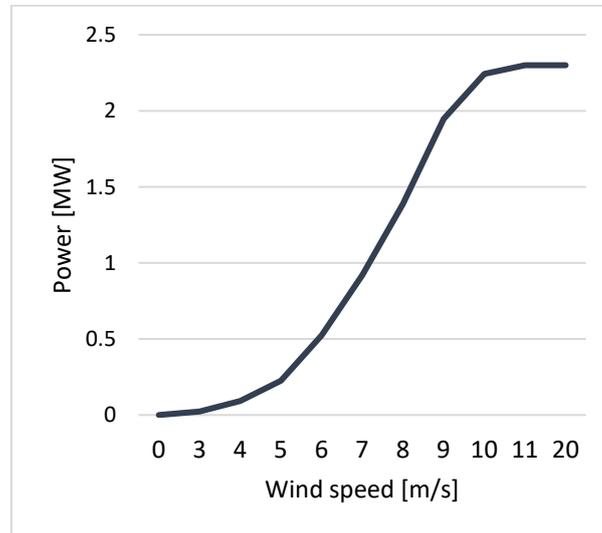

**Fig. 3.** Power curve of Siemens SWT-2.3-108 wind power plant in Atlite. Used for obtaining time series data on power generation potentials for the eligible land in the cut-out of the DRB countries.

Based on power potentials obtained from Atlite for wind power, annual, nationally averaged CFs ranged from 3.3% to 7.8% depending on the country and land layers selected. The numbers cited here were considered small compared to the wind power facilities operating in the DRB countries. This reflects poor data quality rather than the region's absence of wind potential. A geospatial resolution of 30x30 km is insufficient when estimating areas with potentially high wind potentials that can be observed in smaller areas. Additionally, the ERA5 does not apply correction factors to the wind data. As such, we use a separate approach outlined in Section 3.2.3. Using Atlite with the ERA5 dataset provides the hourly data needed for the model. Underrepresentation of areas with high resource availability is not present in the case of solar power; see Fig. 4, since the irradiation is less impacted by altitude, terrain, and location than wind power.

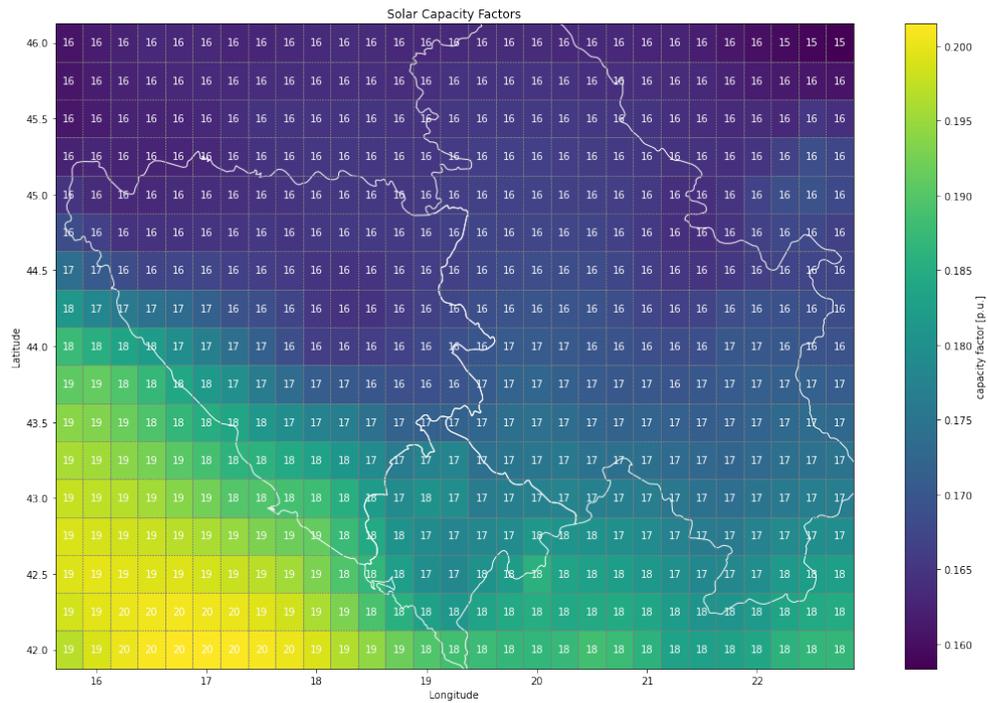

**Fig. 4.** Map of solar capacity factors for a fixed panel by grid cell for the DRB countries, generated using Atlite.

3.2.3   Wind power generation potentials using GWA

To improve the wind power potential estimate for this study, we used the Global Wind Atlas (GWA) version 3 [50]. The GWA is derived from ERA5 reanalysis but uses downscaling processes, resulting in a final resolution of 250m considering local topography and terrain features. The tool provides mean CFs for three different turbine classes. The classes available include the IEC1, IEC2, and IEC3 categories for the 2008-2017 period. The IEC classes correspond to a 100m hub height and rotor diameters of 112, 126 and 136m, respectively. When compared to existing wind generation in the regions, the CF estimates from the GWA are more accurate than those from Atlite.

Annual capacity factors from the GWA used in this study are shown in Fig. 5. For computational considerations, we omit land areas with unfavourable wind conditions, i.e. where wind CFs are less than 10%.

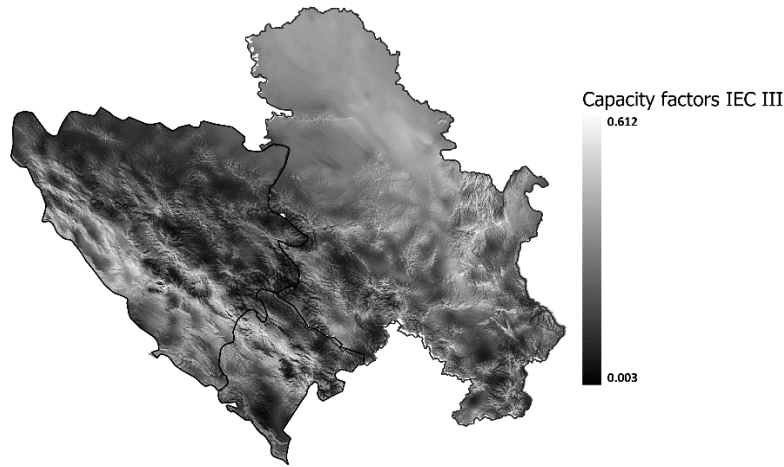

**Fig. 5.** Capacity Factors for IEC Class III wind power plants for the DRB countries from GWA.

The output from the VRE characterisation shown in Fig. 4 and Fig. 5 comprises geospatially explicit CFs for solar and wind power. We also obtain time-series values of capacity factors from Atlite. These outputs are used as inputs to represent wind and solar power in the OSeMOSYS energy model.

In OSeMOSYS, we define technologies with specific techno-economic characteristics, including total capacity potentials and time-series values for CFs. We assume four CF ranges to represent the wind power in OSeMOSYS in a computationally manageable way while covering the broad range of CFs shown in Fig. 5. These are 10-20%, 20-30%, 30-40%, and 40% or higher. We use these ranges to calculate the average CF for each CF range and country. Using the average CF of each range, we scale the time-series values obtained from Atlite to match the four new averages. These adjusted time-series CFs are then assigned to four wind technology representations in each country in the OSeMOSYS model.

After calculating the different time-series CFs, we calculate the share of the DRB countries' areas, which the four selected CF ranges occupy. Using Python, we perform stratified sampling on the output shown in Fig. 5 using 10,000 points per grid cell for all grid cells shown in Fig. 2. With the shares, we calculated the land availability for wind power for each CF range and unit of eligible land from Fig. 2 according to Eq. 1.

$$\text{Available land}_{\text{CF range, country}} = \text{eligible land}_{\text{country}} * \text{share}_{\text{CF range}}$$

**Eq. 1.** Calculating available land for a given CF range in each country

We do not restrict investments for VRE technologies competing for the available land. The model can invest fully in wind, solar, or a mix of both technologies. Investments in a particular land area are based on the techno-economic characteristics of the power plant types. To differentiate between wind technologies within the CF ranges in the model, we use technology names corresponding to the technology type, i.e. wind and the CF range. The solar power resource is divided into multiple technologies to keep the names unique and to pair them with the different wind technologies in user-defined constraints. Each pair of wind and solar technologies for a given CF range, e.g. 30%, is then coupled with a land technology with an availability corresponding to the output from Eq. 1 for that particular CF range. Based on the assumption that the VRE technologies have an area-specific installable capacity of 1.7 MW/km$^2$, we add a land use of 0.588 km2 per MW installed capacity. This allows the model to invest in wind or solar until the available land corresponding to their CF range is exhausted. When power plant developments have occupied all land within a CF range, there can be no additional capacity additions of wind or solar power within that specific CF range. The VRE technologies considered in the OSeMOSYS model for this analysis are summarized in the supplementary materials.

Using the higher average CFs to normalize the time series wind potential data from Atlite is a simplified way of bridging the gap between the two data sets. What justifies this approach in our case is the nature of the model setup. Existing and planned hydropower or thermal power plants in the model are represented as just one aggregated technology per power generation type, with the HPP cascade and the VRE representation being an exception. By following the above-described methodology, we can include hourly power generation potentials of wind and solar, account for eligible land for VRE deployment in each country, and better represent the CFs for wind power based on CF data from the GWA.

### 3.3 Temporal resolution

Long-term expansion models of the power sector generally consider system developments over several decades, which is a computationally intensive process. A high temporal resolution within the modelling period further increases this computational load. Consequently, power sector and energy models do not typically represent each hour in the year but instead use representative periods (e.g., days) [51, 52]. This section describes the method used to construct representative periods and the temporal representation of VRE availability based on climate data.

#### 3.3.1 Temporal structure

The created OSeMOSYS energy model seeks to optimize power system investment and operational decisions for 2020–2050. Each year is represented by fifteen "representative days", each assigned a weight corresponding to its relative frequency. The motivation for using representative days is to decrease the size and computational requirements of the overall energy system model. Using fifteen representative days at hourly resolution, each model year consists of 15x24 = 360 time steps (TS). By contrast, if no temporal aggregation technique is employed, each model year would consist of 8760 time steps, and the model solve time and memory requirements would be computationally intractable. Fig. 6 shows the load and generation duration curves.

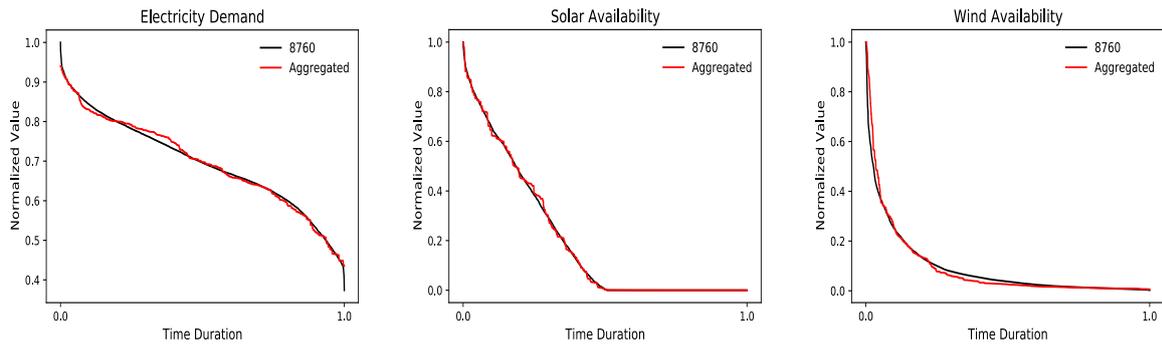

**Fig. 6.** Load and generation duration curves for demand, solar and wind of 15 TS approximation compared to aggregated 8760 TS input data.

The representative days and their respective weights are selected using the agglomerative hierarchical clustering algorithm outlined by Nahmmacher et al. [51] and used more recently in studies by Palmer-Wilson et al. [53] and Keller et al. [54, 55]. Hourly data for electricity demand, wind power availability, and solar power availability across the DRB countries is collected and normalized for a set of historical days[3]. A hierarchical clustering algorithm is then used to group the historical days into 15 clusters, where days within a cluster are broadly similar in load, wind, and solar characteristics. For each cluster, the day closest to the cluster's centroid is selected as the representative day and is assigned a weight proportional to the cluster's relative size. Finally, each representative day's load, wind, and solar time series are scaled to match correct annual averages. Fig. 6 shows the fitment of the 15 TS approximation compared to the input data. A more detailed overview of the methodology can be found in [51].

3.3.2   Reference energy system

Fig. 7 illustrates the Reference Energy System (REF). The REF is a network representation of technical activities required to supply electricity to meet the final demand. It shows the connections between supply and demand, including land and water resources, in the OSeMOSYS energy model. Fossil fuels considered are coal and natural gas. Fossil fuel resources are consumed by coal and gas power plants in proportion to their power generation output. The land resource obtained using Atlite and the CLC is fed into the model as an input. The land is utilized by wind and solar technologies in proportion to the increase in installed capacity, representing land use for constructing energy infrastructure. Water availability for HPPs in the DRB is considered an input that constrains the hydropower cascade. Power imports and exports include cross-border power exchanges between the countries of the DRB, as well as from adjacent countries, including Croatia, Hungary, and Italy, among others. For each country, the resources, power generation technologies, and losses related to transmission and distribution are separately accounted for.

---

[3] Here we are using 2020 data. Wind and solar availability data is collected following the approach outlined in Section 4.1.1 and historical electricity demand data is collected from ENTSO-E.

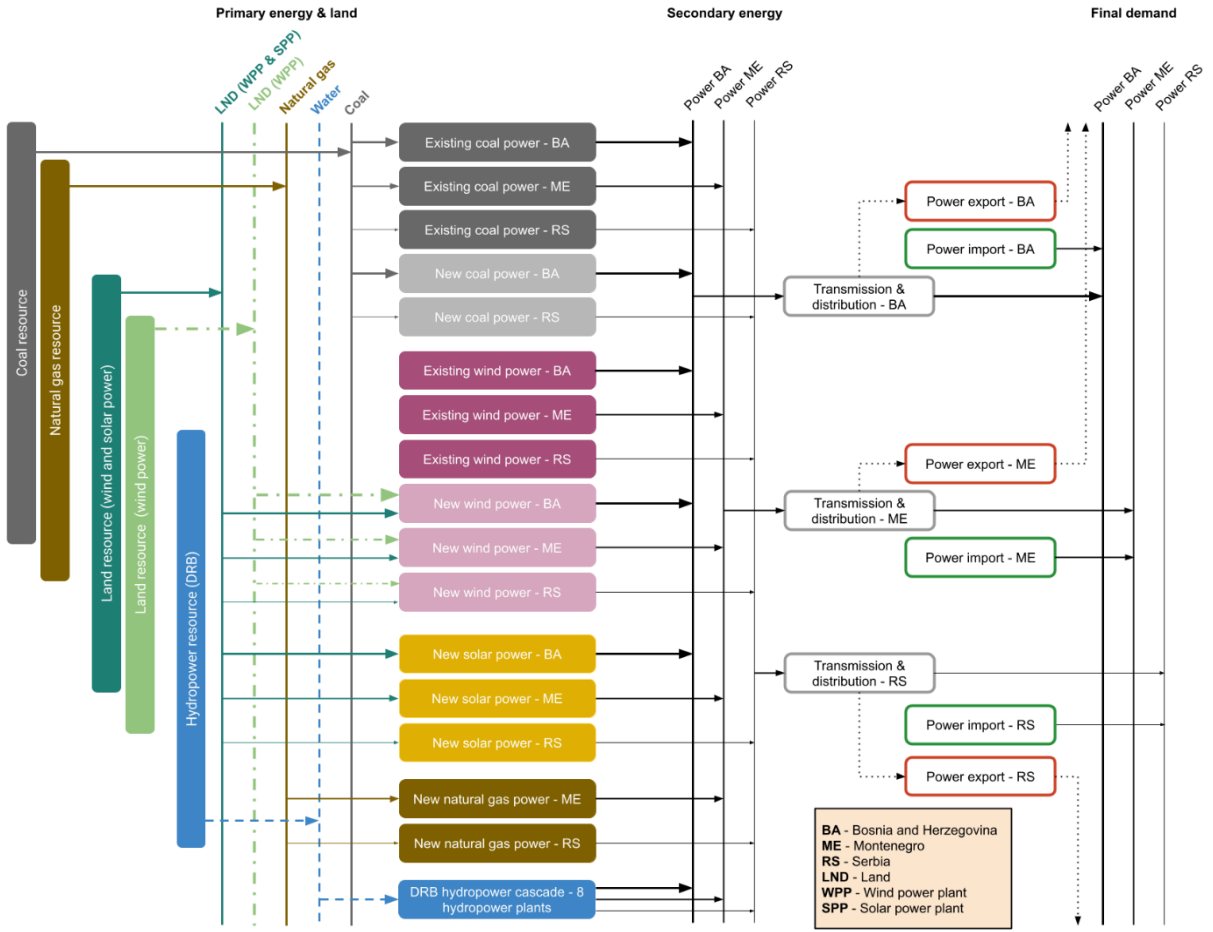

**Fig. 7.** Illustration of the reference energy system for each of the DRB countries, including demands, trade, and power sector infrastructure

### 3.3.3 Hydropower cascade

Fig. 8 illustrates in more detail the DRB hydropower cascade depicted in blue in Fig. 7. We derive data for the cascade water inputs from the HypeWeb model, more specifically, the HYPE model for Europe (E-HYPE) [56]. From the E-HYPE 3.1.1 model version, we calculate the average daily river discharge during the 1981-2010 period for each of the following rivers: Ćehotina, Lim, Piva, Tara, and Uvac. The resolution of this data corresponds to daily flows. This entails that flows within each TS are constant and equal to the daily average based on the E-HYPE model data. Flows change between the 15 selected TS. Water enters the cascade through the upstream river segments and the catchments. River segment capacities and water flows are input parameters. These capacities are fixed to the maximum average flow of each day. The WFDEI dataset of historical precipitation and temperature is used as forcing data in this simulation [56]. The capacities do not vary across the different years in the model.

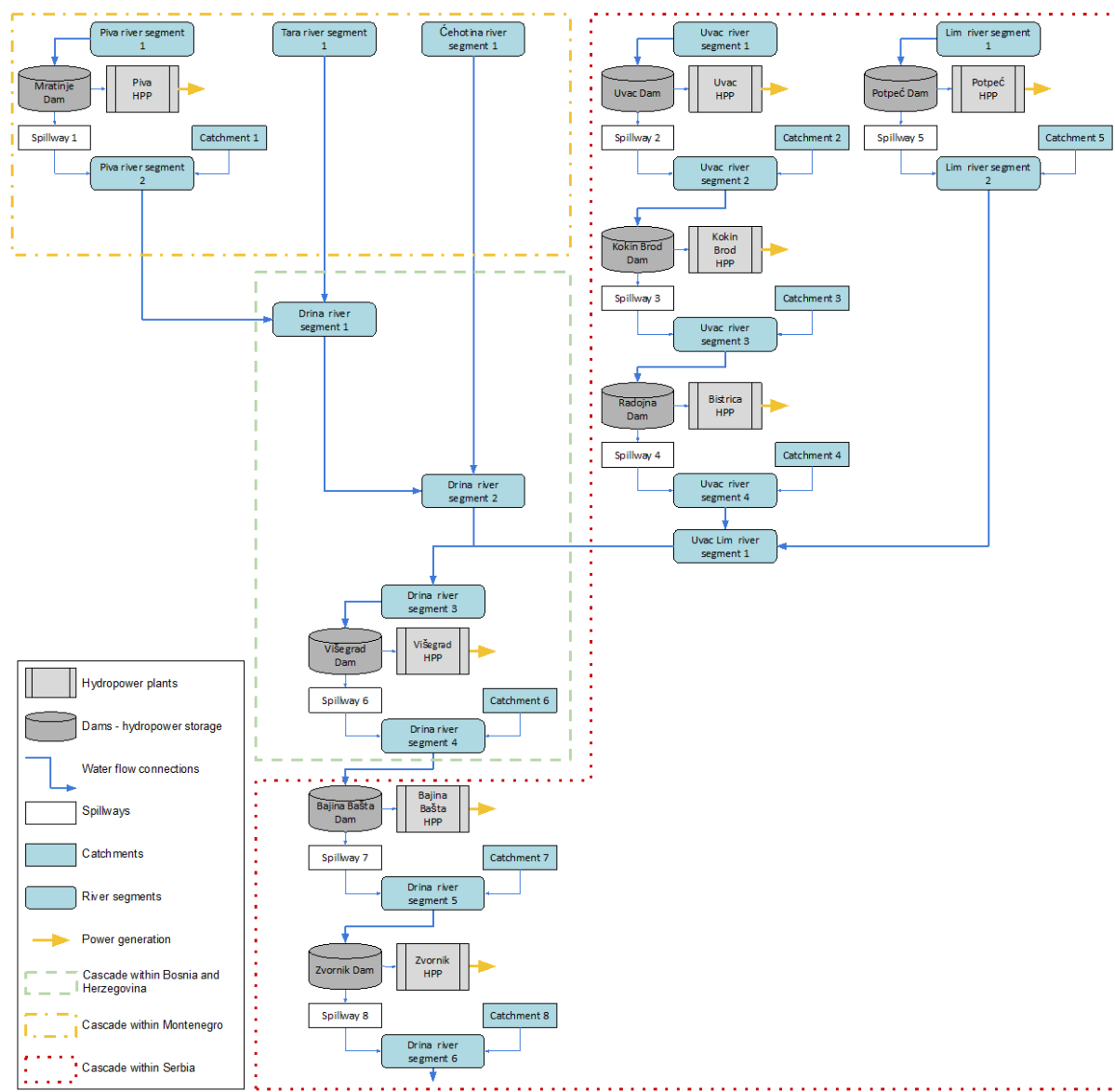

**Fig. 8.** Structure of the HPP cascade. Boxes represent technologies in OSeMOSYS while connecting lines represent commodities. In this figure, the commodities are water flowing between the power plants, spillways, dams, catchments, and river

segments. Catchments in this representation are aggregations of small tributaries or streams entering the DRB cascade.

### 3.4 Explored development pathways – a scenario analysis.

We created three scenarios to explore potential development pathways for the power sectors of the DRB countries. A Reference scenario is presented first to represent a baseline, followed by an Emission Limit Scenario to explore achieving net-zero emissions by 2050. Finally, with the Agricultural land for wind power scenario, we examine wind power development's effect on the DRB countries' vast agricultural land could have on the power sector. The three scenarios are explored with two variations: high- and low-cost trade alternatives. For the low-cost variation, we used average wholesale electricity prices for the period 2018-2020 based on seven selected spot markets (EPEX Germany, EPEX Austria, SIPX, HUPXDAM, OPCOM, SEEPEX, and CROPEX) [57]. For the high-cost variations, we used the values for 2021. We made this differentiation given Europe's fossil gas price surge in 2021, which more than doubled wholesale electricity prices.

**Reference scenario (REF)**. The reference scenario serves as the baseline for the scenario analysis. All other scenarios are compared with results from this baseline. The model may invest in technologies currently employed in the power sectors of the DRB countries. These include coal, solar, hydro, and wind power, with several exceptions. As the government of Montenegro has already cancelled its last coal project and is presently focused on accelerating the retirement of its remaining CFTPP [58], no new CFTPP projects are permitted in Montenegro in this scenario. Since projects concerning planned HPPs in the DRB are uncertain, the hydropower cascade is not expanded in this scenario. This scenario aims to provide insights into the development of the power sector based on current techno-economic parameters obtained from literature and consultations with local stakeholders without any additional policy measures. The global assumptions are consistent across all scenarios and are provided in the supplementary materials.

**Agricultural land for wind power scenario (AG)**. Here, we relax the upper bound on wind capacity by making wind development possible on agricultural land. As a percentage of their total land area, Bosnia and Herzegovina and Montenegro have a modest 12% share of agricultural land, while Serbia has 40%. This scenario aims to inform us of the role that wind power plants (WPPs) on agricultural land may play in decarbonizing power sectors in the DRB countries.

**Emission limit scenario (EL)**. Assuming future EU integration of the DRB countries, this scenario aims to provide insight into the development of their power sectors if emissions are restricted to the EU's 2030 and 2050 GHG reduction targets. Compared to 1990 emissions levels, the target values correspond to a 55% reduction by 2030 and a net-zero emission level by 2050. The applied emission limit is shown in the supplementary materials.

# 4   Results

In this section, we present the key findings of the analysis. The results are reported in an aggregate form, combining the findings for all three DRB countries. We present the estimates for VRE potentials, their role in supporting the transition to net zero by 2050, and the impact new VRE developments can have on the existing hydropower cascade regarding power generation and cost competitiveness.

### 4.1   VRE Potential

Based on the methodology described in the section 3.2, the findings reported in Table 2 indicate a combined wind and solar potential of 94.4 GW. Most of the VRE potential in Bosnia, Herzegovina, and Montenegro is not located on agricultural land. In contrast, wind potentials on agricultural land in Serbia account for over 80% of its total VRE potential. About half of the wind potential on agricultural land in Bosnia and Herzegovina and Montenegro has CFs within the 15% range. In contrast, most of Serbia's CFs on the same land type are around 35%.

**Table 2.** Wind and solar power potentials of the DRB countries

| Country | Average CF range wind [%] | Wind or solar power potential [GW] on shared land areas | Wind power potential on agricultural land [GW] | Solar power potential [GW] on land with low CF for wind power (<10%) |
|---|---|---|---|---|
| **Bosnia and Herzegovina** | | | | |
| | 15.6 | 2.7 | 4.2 | 2.1 |
| | 25 | 3.8 | 2.9 | - |
| | 34.8 | 3.8 | 0.9 | - |
| | 45.3 | 2.3 | 0.4 | - |
| *Subtotal* | - | *12.6* | *8.4* | *2.1* |
| **Montenegro** | | | | |
| | 15.2 | 2.1 | 0.8 | 1.8 |
| | 24.8 | 2.2 | 0.6 | - |
| | 34.5 | 1.5 | 0.4 | - |
| | 45 | 0.7 | 0.2 | - |
| *Subtotal* | - | *6.5* | *2* | *1.8* |
| **Serbia** | | | | |
| | 15.9 | 2.5 | 5.2 | 1.6 |
| | 24.9 | 3.9 | 15.4 | - |
| | 34.8 | 3.4 | 26.7 | - |
| | 43.8 | 0.7 | 1.6 | - |
| *Subtotal* | - | *10.5* | *48.9* | *1.6* |
| **Total** | - | **29.6** | **59.3** | **5.5** |

## 4.2 Impact of VRE on the hydropower cascade

Fig. 9 illustrates the water levels in the HPP cascade for the low-cost trade alternative REF scenario. On the y-axis, the storage level is expressed in million cubic meters (MCM), while the x-axis represents the 360 time steps. The dark blue line represents the average storage level for each time step over the 2020–2050 period. With a 95% confidence interval, the bright blue areas surrounding the mean indicate the minimum and maximum values.

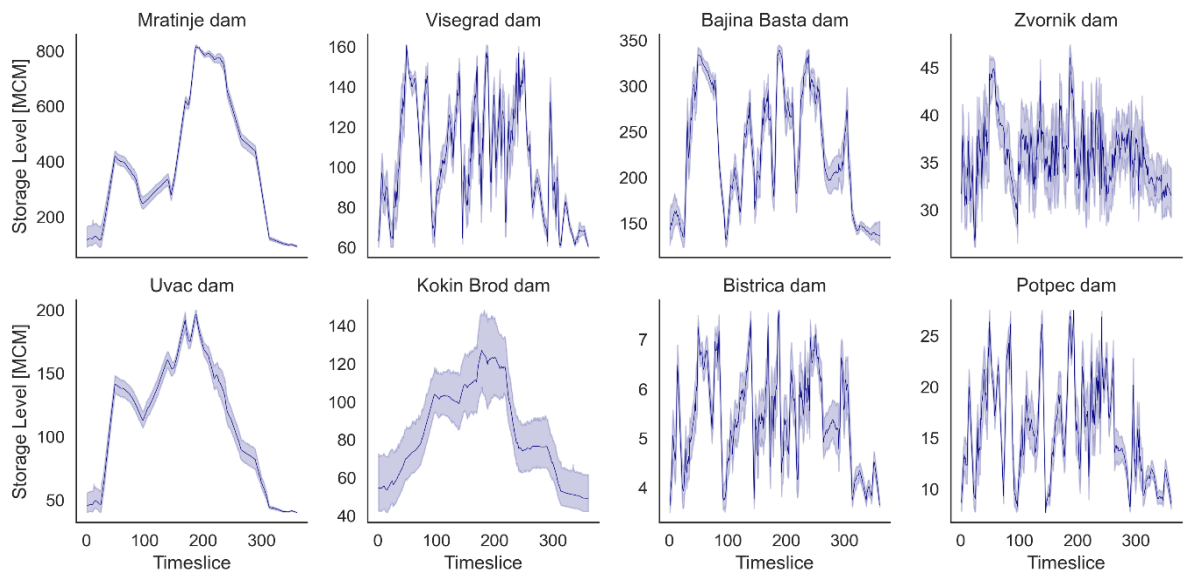

**Fig. 9.** HPP cascade water levels for the REF scenario in each time step for the 2020–2050 period

Assuming high-cost trade, investments in solar increase by 45%, or 2.4 GW, compared to the low-cost trade alternative REF scenario. These capacity additions result in increased power generation from solar, as shown in Fig. 10 (b). The added solar capacity drives an increase in generation from the cascade during the modelling period. The model utilizes a combination of solar and CFTPP generation based on their power supply profiles, reducing wind and hydropower generation outside the basin.

With increased solar power generation in the high-cost REF scenario, the hydropower cascade, as the only storage option in the model, enables a more balanced solar power output by generating 35 GWh less power in 2050 relative to the low-cost REF scenario, which is characterized by comparatively fewer solar PV additions. Meanwhile, during the 183 hours in 2050 when solar is unavailable, the cascade generates 138 GWh more power in the high-cost REF scenario than in the low-cost REF scenario.

Additionally, Fig. 10 (a) indicates import dependence during time steps with low VRE availability. Changing power import and export prices alters the investments and operational strategy. The change is considerable, with imports decreasing under the high-cost REF scenario while investments in VRE technologies and CFTPPs increase by 3.1 and 1.6 GW, respectively.

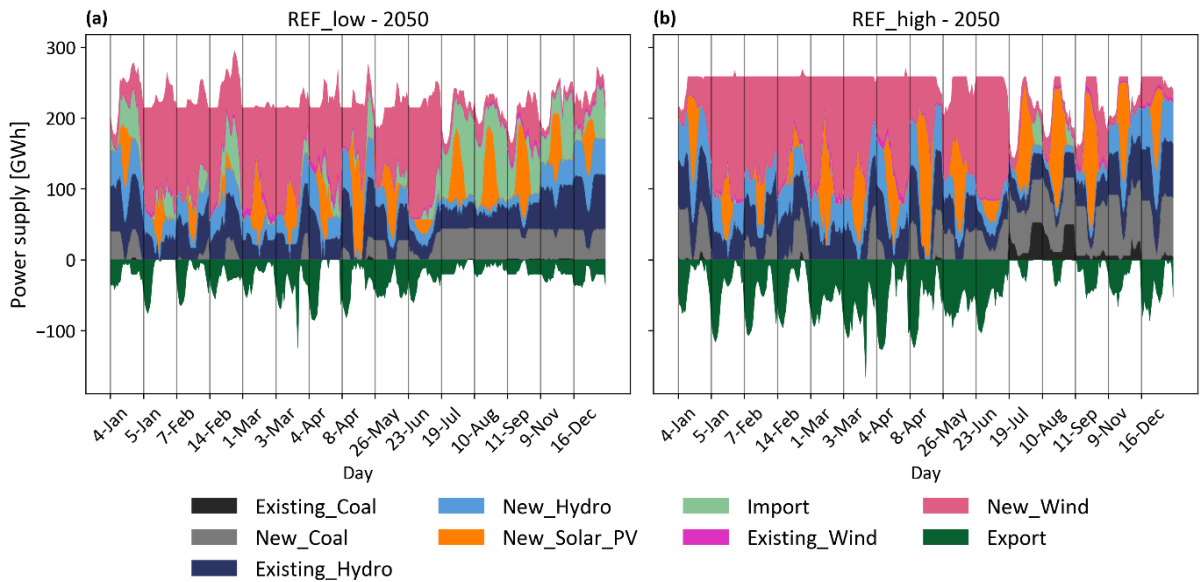

**Fig. 10**. Daily power supply for each of the 15 representative days, based on the REF scenario with different power trade costs in 2050.

### 4.3 Capacity additions to the power sectors of the DRB countries

Fig. 11(a) shows the capacity additions under the REF scenario with low-cost trade. VRE investments correspond to 11.8 GW by 2050, or 68% of all new capacity additions. Capacity additions under the REF scenario with high costs are shown in Fig. 11(b), corresponding to 14.9 GW or 66% of the total new capacity. High trade costs favour power generation capacity investments over imports, as highlighted in section 4.2, explaining the increased capacity additions between Fig. 11 (a) and (b). We observe further VRE additions in the EL scenario, 16.5 and 14 GW, for the high- and low-cost alternatives.

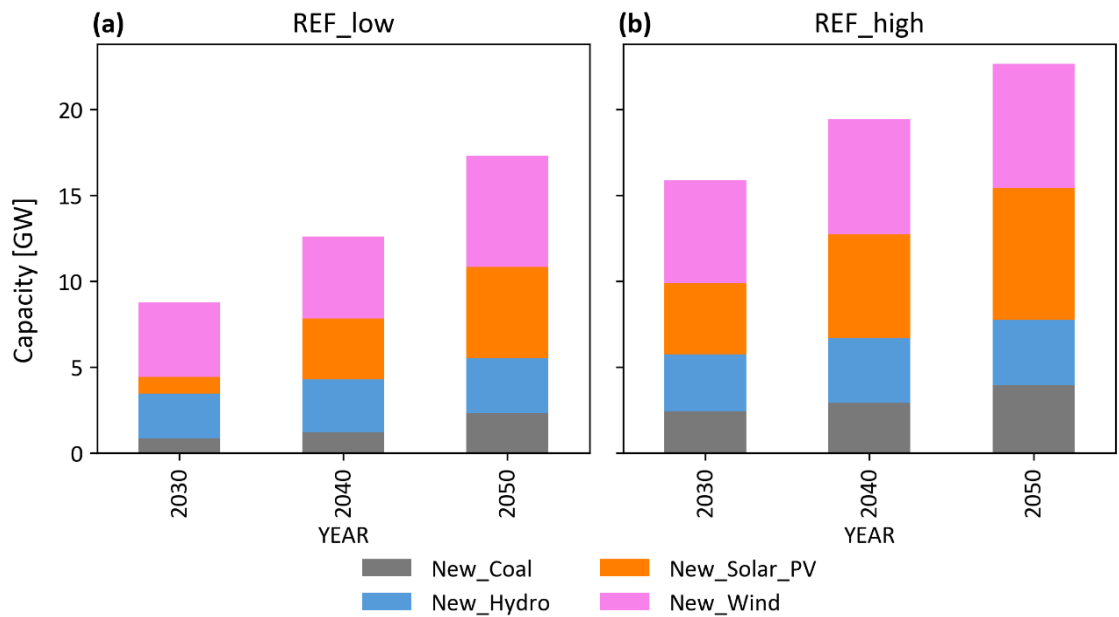

**Fig. 11.** New cumulative capacity additions under the high- and low-cost trade alternatives of the REF scenario.

Fig. 12 shows the difference in capacity additions for the AG and EL scenarios compared to the REF scenario. It includes differences for each scenario's high- and low-cost trade variation. Subplots (a) and (b) show decreased solar investments when additional land available for wind expansion on agricultural land is presented in the model. The increase in wind also includes wind potentials with higher capacities; hence, subplot (c) only shows an overall capacity decrease in 2040 compared to the RES. Similarly, the decommissioning of CFTPPs requires a substantial expansion of wind and hydropower capacity, as shown in Fig. 12 (c) and (d) in 2050. Given different generation profiles of supply technologies, solar investments in the EL scenario are lower due to the absence of CFTPP capacities that complement solar in the REF scenario.

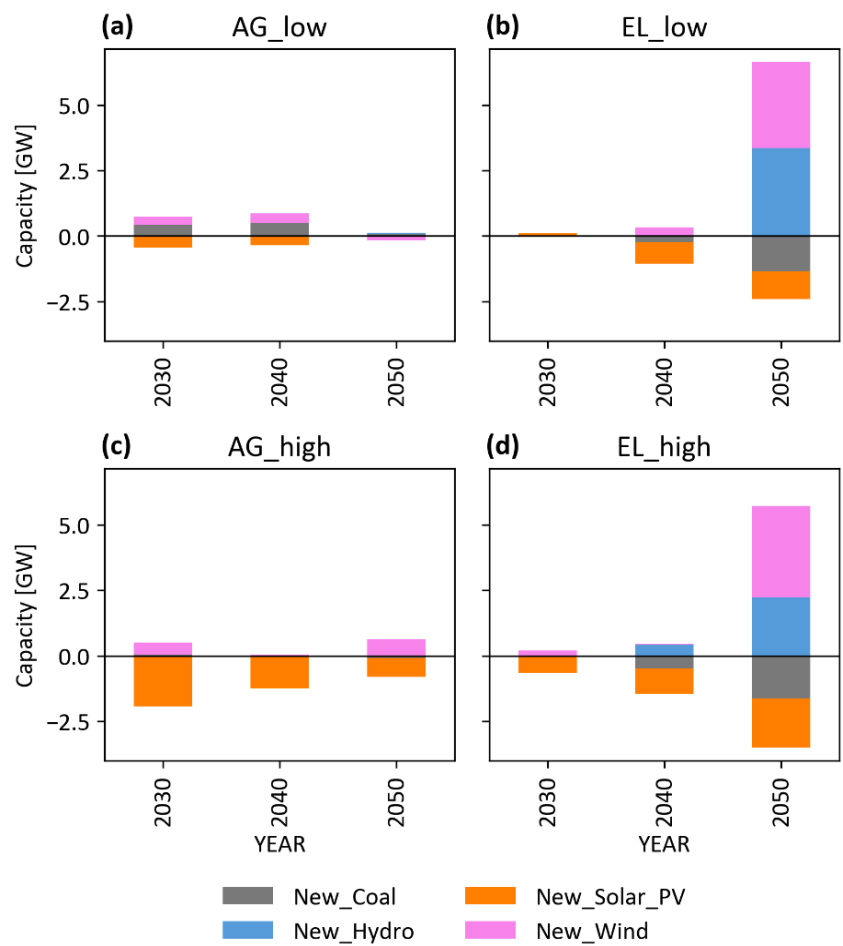

**Fig. 12.** Difference graph of cumulative capacity expansions between the explored scenarios and the REF scenario with high- and low-cost trade. Values given in GW; negative values indicate lower capacity than the REF scenario.

### 4.4 Developments of the power supply across explored scenarios for the DRB countries

Fig. 13 shows the power supply and the power sector expansion across the scenarios. The subplots (a), (b), and (c) show low-cost trade alternatives for the explored scenarios, while (d), (e), and (f) refer to the high-cost alternatives.

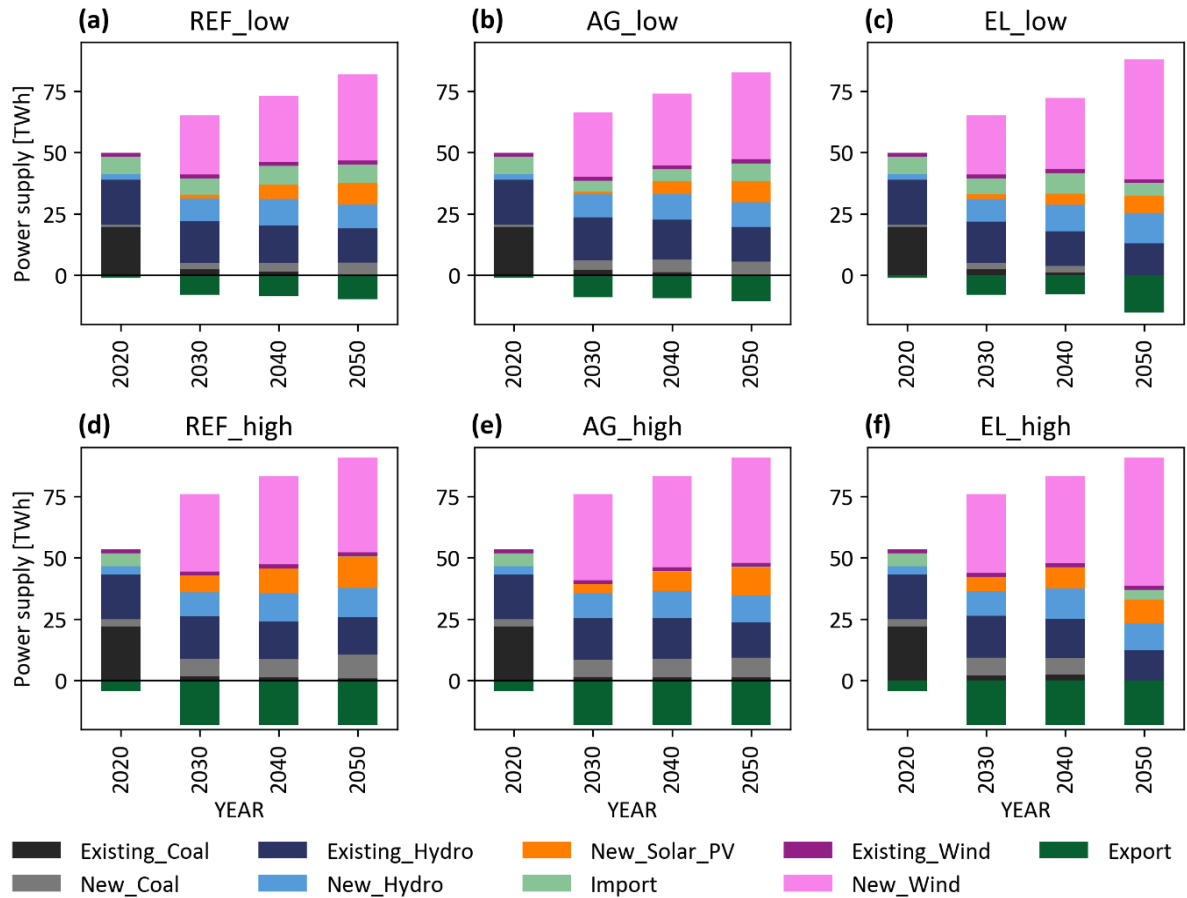

**Fig. 13.** The DRB countries' power supply, imports and exports, all scenarios 2020-2050.

Results shown in Fig. 13 indicate a sharp phase-out of CFTPPs across all scenarios during 2020-2030. The main driver behind this model behaviour is the high costs associated with coal-fired power generation. CFTPPs in the Western Balkans are old and inefficient, and coupled with low coal mine productivity, they lead to high power generation costs, requiring high electricity costs to stay profitable.

Other insights derived from Fig. 13 include that increased power costs incentivize the model to generate more power within the DRB countries to be exported to neighbouring countries. This behaviour is represented in subplots (b), (d), and (f). In addition to more exports, the results show higher shares of CTPPs in these scenarios, which correspond to new, more efficient power plants compared to the current installed CTPP capacities.

Fig. 13 subplots (a), (b), and (c) show scenario outcomes with lower electricity prices. These lead to fewer capacity additions within the DRB countries in favour of imports, lowering the share of CTPPs. The increased import dependence suggests a decrease in the energy security of the analysed countries.

## 4.5 Emissions associated with power generation in the DRB countries

The emissions shown in Fig. 14 represent CO2 emissions from power generation. The emissions include the direct emissions from burning coal for power generation. Overall, the emissions associated with power generation are higher in the high-cost alternative scenarios, as shown in Fig. 14 (b).

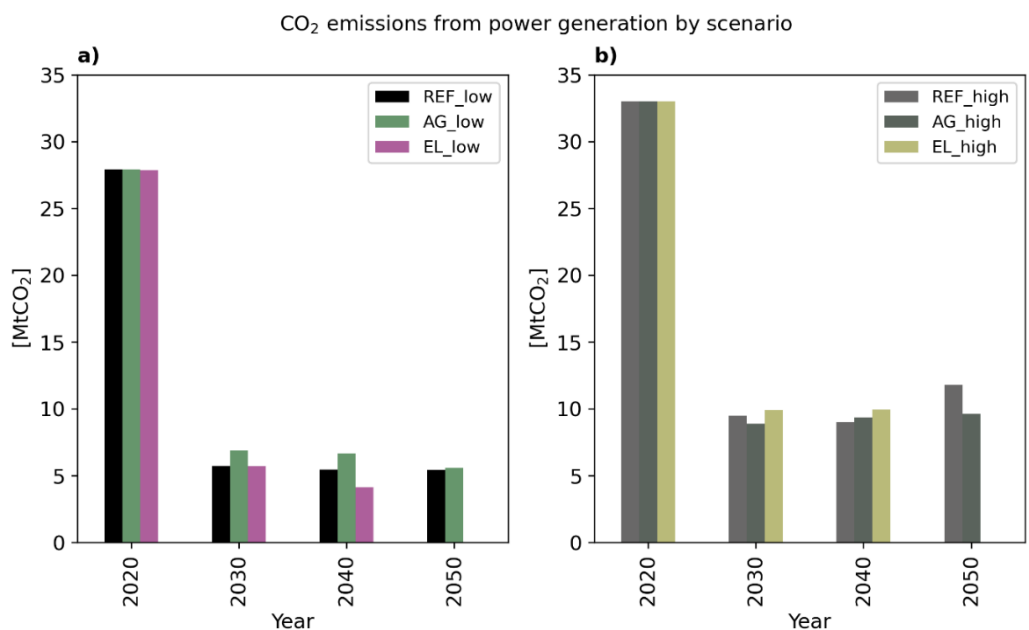

**Fig. 14.** $CO_2$ emissions from power generation by scenario in the DRB countries. Values are provided in $MtCO_2$. Subplot (a) represents the low-cost while (b) represents the high-cost trade alternatives for each scenario.

CO2 emissions shown in Fig. 14 align with the phase-out of CFTPPs shown in Fig. 13. Emissions are higher in the high-cost scenarios due to higher shares of coal in the power supply mix. Power import emissions are not considered in this study. The reason is that the DRB countries export more than they import, and under the assumption that the DRB countries would become EU members, the net emissions from the analysed power sectors would reach net zero from an EU-level perspective. Another reason is that the assumption on the carbon intensity of the neighbouring countries would require modelling efforts outside this paper's scope.

Although a sharp decrease in emissions between 2020 and 2030 is observed, the emission mitigation is still not in line with the latest available scientifically based GHG emission scenarios for climate neutrality of the EU by 2050, which suggest emission reductions of 90 to 95% by 2040 compared to 1990 levels [59].

The investments made in CFTPPs early in the modelling period result in emissions from these power plants until 2050, as shown in Fig. 14. This is due to the long operation life of power generation infrastructure investments.

## 5  Discussion

This section discusses and interprets the results presented in section 4. The findings are discussed in terms of potential implications on the power sector developments and their relation to the purpose of the study and the research questions posed.

One of the aims of this study was to assess the VRE potential within the DRB countries. The results in section 4.1 state that the estimated potential of wind and solar power is 94.4 GW. The breakdown of the total VRE capacity potentials among the DRB countries shows that for Bosnia and Herzegovina, Montenegro, and Serbia, the combined VRE potential is 23.1, 10.3, and 61 GW, respectively. These potentials far exceed the current total installed capacity in 2020 of the DRB countries, which are 4.1, 1, and 7.4 GW, respectively [29, 30, 60]. Current installed capacities of VRE in the DRB countries are less than 1 GW as of 2022 [29, 30, 60]. Compared to earlier assessments from IRENA [11] of VRE potentials shown in section 1, the results of this study show VRE capacity estimates, which are 68% higher.

The results suggest a wind potential of 8.5 GW in Montenegro, assuming all available land, including agricultural land, is utilized for this purpose. This figure far exceeds earlier estimates of 400 MW made by [6-9]. Previous estimates from IRENA [11] stated a technical potential of wind power close to 3 GW in Montenegro, considerably greater compared to earlier studies [9], but less than the 8.5 GW estimate presented in this study. This shows that the capacity estimation of wind potentials has increased over time, which is to be expected given the rapid development of wind turbines.

Section 4.3 presented the capacity additions under the REF scenario and compared those to the capacity expansion in the AG and EL scenarios. The cost-optimization and scenario analysis results suggest investments in wind and solar power corresponding to 10.7 and 5.8 GW, respectively, under the assumption of net zero by 2050 and a high trade cost. This VRE capacity addition corresponds to 17.5% of the total VRE potential estimated for the DRB countries.

Wind power investments in the model are within the 25-45% range. Our results align with observed values from investments in the United States in 2020, where the minimum and maximum CF for wind power plants built in 2019 were 24 and 56%, respectively [61]. Thus, we expect the VRE potential to be sufficient to support the power sector decarbonization of the DRB countries. It should be noted that wind potentials with lower CF ranges are not cost-competitive compared to hydropower and imports across the scenarios. As nearly two-thirds of the VRE potentials are on agricultural land, governments must develop policies allowing WPP deployment in these areas.

VRE penetration by 2050 across the analysed scenarios ranges from 65-68 %. Our general upscaling of VRE in the power supply seems, although unprecedented in the DRB countries context, reasonable. Please refer to [62-65] showing the plausibility of high VRE shares in the power supply mix.

VRE constitutes nearly 70 % of all new capacity additions, as shown in section 4.1. This leads to increased power shares from VRE sources, with wind showing the most rapid growth. Interestingly, these findings align with the developments in the power sectors in the DRB countries during the last few years. However, utility-scale solar power developments have not gained traction in the region. The model results for the REF scenario support this investment trend under the low-cost trade alternative, where no solar capacity additions are observed until 2030. By contrast, under the high-cost EL scenario, investments in solar are observed as early as 2021.

Power generation from coal in the DRB countries is subsidized through direct and indirect subsidies. The public sector and state-owned enterprises provide direct subsidies through investments and fiscal support, including loans and guarantees. These have for the 2015-2017 period been on average 4.12 EUR/MWh in Bosnia and Herzegovina [66], while the direct subsidies in Montenegro and Serbia were 0.7 and 3.87 EUR/MWh, respectively [67]. The indirect subsidies include non-payment of carbon dioxide emissions.

The cost of power generation from coal in the DRB countries is around 70 EUR/MWh. This figure is much higher compared to the end-user prices, which in Bosnia and Herzegovina is 34.2 EUR/MWh for households and 39.8 EUR/MWh for industry [66]. Since state-owned enterprises control most of the power-generating capacities in the DRB countries, they obtain lower power generation costs on average when both coal and hydropower are jointly considered. The model does not capture this behaviour, as it optimizes based on cost and does not consider other factors, such as job security for the mining sector or employment in the coal-fired power generation supply chain. Instead, the model shows the power supply and capacity additions based on the cheapest alternative to meet the given demand at any time in the modelling period.

In addition to replacing thermal capacity, the model invests in both wind and hydropower since their availability profiles are complementary. We observe this dynamic in all graphs in Fig. 12, but most clearly in Fig. 12 (d), illustrating net-zero emissions through the removal of coal from the power supply. In some of the 15 representative days, the capacity factors of VRE are low, while demand is high. The model, therefore, relies on power imports to meet the demand when electricity prices are low. The high cost of electricity imports and exports drives the model to invest in additional capacity to reduce import dependence while exporting with high profits.

Wholesale electricity prices significantly have a significant impact on the power sector developments. Fig. 13 indicates an accelerated investment rate in solar PV, increased exports from the DRB countries, and more investments in CFTPPs compared to low-cost trade scenarios.

The DRB countries can reduce their vulnerability to imports at high prices by expanding their capacity of VRE technologies and hydropower. Additionally, excess power generation from these sources in times of low domestic demand can be used for exports at a higher cost to neighbouring EU countries, such as Italy, Hungary, and Greece. Moreover, the potential introduction of the Carbon Border Adjustment Mechanism (CBAM) in the EU is another reason for increased investments in renewables. The CBAM would entail additional taxation on power exported to the EU from the Western Balkan countries. This could result in new CFTPPs becoming stranded assets since the CBAM taxation would further compromise their cost competitiveness. The model results confirm the fact that the DRB countries are net exporters. However, introducing the CBAM could reduce power export revenues from the DRB countries, making them more import-dependent and with less funds to invest in expanding and maintaining their power sectors. The findings suggest accelerated developments of VRE projects in the DRB countries so that domestic demand can be met with a limited need for electricity imports. Additionally, it calls for the implementation of ETS, which, in contrast to the CBAM, would provide the DRB countries with financial means to invest in more renewable energy.

From Fig. 14 we observe that a higher cost of trade results in increased emissions. This is in line with findings from Fig. 13, which shows the continued use of CFTPPs under the high-cost trade alternative scenarios. The model findings suggest that the expansion of renewables in favour of CFTPPs is the main driving factor of the $CO_2$ mitigation observed in Fig. 14.

Since the time steps are not sequential in the model, we cannot assess the effect of seasonal variations of water availability on storage levels under the explored scenarios. We can, however, compare the total water levels in each dam between different scenarios, the power output of the HPPs, and their operational rules. The results suggest that scenarios with higher shares of SPPs and CFTPPs utilize the HPP cascade for power generation more than scenarios where WPPs and HPPs outside the basin are the main capacity additions.

The operation of the cascade HPPs shifts to increased short-term balancing of renewables by the cascade, moving from baseload to more responsive power generation patterns. The finding highlights that potential expansion of the HPP cascade can enable larger shares of solar power, resulting in high shares of renewables in the power supply coupled with the balancing capabilities of hydropower.

# 6  Conclusions

Having among the largest shares of coal-based power generation in Europe, the DRB countries of Bosnia and Herzegovina, Montenegro, and Serbia must take action to meet the EU's goal of net zero emissions by 2050. In this paper, we created a power sector model for the DRB countries with a scenario analysis exploring different development pathways.

We present a novel approach for assessing the VRE resource potentials by combining time-series data on availabilities from Atlite with high-resolution capacity factor data obtained from the GWA. The findings indicate a capacity potential of 94.4 GW of VRE in the DRB countries, of which 59.3 GW or 63% relate to wind power. The current installed capacity within the DRB countries is 12.5 GW, of which 627 MW is wind power. This suggests that the VRE potential is largely untapped. Results from this study suggest a 68% increase in VRE potential estimates compared to previous assessments.

Findings from the EL scenario, where net-zero emissions are expected by 2050, show wind and solar power investments corresponding to 10.7 and 5.8 GW, respectively. These constitute 17% of the assessed VRE potential presented in this study. We thus conclude that regional VRE potentials are sufficient to support the transition of the DRB power sectors to net zero by 2050. VRE technologies are crucial in mitigating power sector emissions, and their shares in new capacity additions range from 65-68% in the explored scenarios.

VRE expansion in the DRB countries has a limited effect on power generation from the HPP cascade. However, the results indicate increased power generation from the cascade under the high-cost REF scenario, where higher shares of solar power are present. As more solar is deployed in the system, the HPP cascade increasingly acts as a short-term balancing option, moving from baseload to more responsive power generation patterns.

VRE technologies are crucial for power supply to meet demand while adhering to emission reduction targets. Delaying renewable energy developments could lead to stranded assets after the introduction of the CBAM. Undercapacity could be costly, especially given the current costs of cross-border European power trade and the risk of reduced import availability from EU countries surrounding the Western Balkans. Additionally, not aligning with the Sofia Declaration commitments might hinder EU accession.

# 7   Limitations and Future Research

In this section, we highlight the limitations of this paper, including potential topics of future research needs that could expand the work presented in this paper.

The land availability assessment for VRE developments was limited to utility-scale technology options. It did not include rooftop solar, which could be utilized in urban settings, nor solar on agricultural land. Since costs relating to the expansion of transmission and distribution lines, distances from the grid, slopes or difficult-to-reach areas were not included when assessing the VRE capacity potential, the total calculated VRE capacity presented in this paper may be less utilized than the results suggest. In contrast, improvements in efficiency and capacity factors of VRE technologies, which are likely to improve their cost-competitiveness, are not included. Considering that the model developed for DRB countries is intended to inform long-term energy infrastructure investments and not site-specific power generation projects, future research could combine the presented methodology for estimating VRE potentials with site-specific analyses.

We acknowledge that introducing different storage alternatives could affect the relationship between the power supply options. It would, for example, enable excess solar to be stored in pumped hydro storage (PHS) or batteries. Adding this flexibility would remove the model's need to couple power supply alternatives solely by their availability. We thus expect the investments to be more flexible than the results suggest in Fig. 12.

In the model, the power demand is a driving factor for the expansion of the power sector. We use data from the current demand profiles and demand projections based on projections made by the local transmission system operators. An interesting point to consider going forward is the impact of demand reductions based on energy efficiency measures. Energy efficiency improvements in the Western Balkans have, over the past decade, been the basis for financial support in the region from the Regional Energy Efficiency Programme launched in 2012 [68], to the Energy Support Package [69], a proposal comprising 1 billion Euro, was put forward in 2022 to promote diversity of energy supplies, increase renewable energy, and improve energy efficiency.

Cross-border trade is present in all explored scenarios. As highlighted in this paper, the DRB countries could become increasingly import-dependent in case of power shortages due to decommissioning of thermal power or delayed investments in low-carbon technologies. Given disruptions in the European power markets caused by the Ukraine conflict, imports could be more expensive and unavailable. The impact of import availability on the expansion of the DRB countries' power sector may be better understood by further modelling the region with different levels of import availability.

## Declaration of Competing Interests

The authors declare that they have no known competing financial interests or personal relationships that could have appeared to influence the work reported in this paper.

## Acknowledgement

This project has received funding from the European Union's Horizon 2020 research and innovation programme under grant agreement No 101022622. The authors would like to acknowledge the United Nations Economic Commission for Europe (UNECE) for the funding provided and for facilitating access to stakeholders, together with Global Water Partnership Mediterranean (GWP-Med) through the project "Promoting the Sustainable Management of Natural Resources in Southeastern Europe, through the use of Nexus approach (2016-2022)" funded by the Austrian Development Agency. The authors thank Francesco Gardumi and Youssef Almulla for contributing to the hydropower cascade development.

## CRediT author statement

**Emir Fejzic**: conceptualization, methodology, writing of the original draft, reviewing and editing, software, visualization and data curation. **Taco Niet**: conceptualization, methodology, writing - review and editing, supervision, and funding acquisition. **Cameron Wade**: methodology, writing - review and editing, software, and visualization. **Will Usher**: conceptualization, writing -review and editing, supervision, and funding acquisition.

## Supplementary material

The supplementary material to this paper is available at the following [Zenodo](#) deposit.

## Data availability

Datasets related to this article can be found at doi: 10.5281/zenodo.7895232.